\documentstyle[12pt,aaspp4]{article}  
  
\begin{document}  
 
\begin {flushright} 
OU-TAP-36 \\ 
May 1996 
\end{flushright} 
 
\title{\bf SCALE-INVARIANT CORRELATION FUNCTIONS OF COSMOLOGICAL DENSITY  
FLUCTUATIONS IN THE STRONG CLUSTERING REGIME}  
  
\author{Taihei Yano and Naoteru Gouda}  
  
\affil{Department of Earth and Space Science,  
Graduate School of Science, Osaka University  
Toyonaka, Osaka 560, Japan\\  
E-mail: yano, gouda@vega.ess.sci.osaka-u.ac.jp}  
  
\begin{abstract}  
We have investigated the scale-invariant solutions of the BBGKY equations  
for spatial correlation functions of cosmological density fluctuations   
and the mean relative peculiar velocity in the strongly nonlinear regime.  
  
It is found that the solutions for the mean relative physical velocity depend on the three-point spatial correlation function and the skewness of the velocity fields.  We find that the stable condition in which  the mean relative physical velocity vanishes on the virialized regions is satisfied only under the assumptions which Davis \& Peebles took in his paper. It is found, however, that their assumptions may not be general in real.  
  
The power index of the two-point correlation function in the strongly nonlinear regime depends on the mean  
relative peculiar velocity, the three-point correlation function and the skewness. If the self-similar solutions exist, the power index in the strongly nonlinear regime is related to the power index of the initial power spectrum and its relation depends on the three-point correlation function and the skewness through the mean relative peculiar velocity. 
  
Furthermore it is found that the mean relative physical velocity should have the values between 0 and the Hubble expanding velocity. When the mean relative  
physical velocity equals to the Hubble expanding velocity, there might exist  self-similar solutions in which the power index of the two-point correlation function in the strongly nonlinear regime is independent of the initial  
power index $n$.  
\end{abstract}  
{\it Subject headings}:  
cosmology:theory-large scale structures-correlation function   
  
\section{INTRODUCTION}  
  
In our universe, there are many kinds of structures  
 such as galaxies, clusters of galaxies and super-clusters of galaxies.  
 It is not known clearly still today how galaxies and the large-scale structures have been formed after the universe was born.  
This is one of the most important problems in the cosmology.  
  In the standard scenario of the large-scale structure formations,   
it is considered that small fluctuations at the early time   
grow as time increases due to the gravitational instability.   
  Hence it is very important to clarify the evolution of density fluctuations.  
Here we consider the density fluctuations of the collisionless particles such as dark matters because our interest is concentrated on the effect of the self-gravity.  
  When the density fluctuations are much smaller than unity at the early time,  
time evolution of the small density fluctuations   
can be analyzed by making use of the linear theory.  
  In this regime, we can understand analytically how the small fluctuations   
grow(Peebles 1980, 1993).  
  When the amplitude of these small fluctuations becomes as large as unity, that is,   
the fluctuations become to stay at the quasi-linear stage,   
we can not make use of the linear theory.  
In this regime, the higher order  perturbation methods or the  Zel'dovich approximation is often used for the analysis.  
  Moreover the density fluctuations continue to grow as time increases and then the amplitude of these fluctuations  
becomes much larger than unity. At last the caustics of density fields appear anywhere in this regime.  
  In this strongly nonlinear regime, the analytical approach is impossible.  
However the nonlinear phenomena of the self-gravity is very interesting and  
mportant for not only the large-scale structure formations, but also the academic interest on the nonlinear dynamics.  Hence we believe that it is very necessary  
to  
understand clearly the nonlinear behavior of the density fluctuations.  
For example, we are interested in the two-point spatial correlation  
functions in the strongly nonlinear regime.  In this regime, it is found   
{}from the $N$-body simulations that the two-point correlation functions obey the power law.  This result is reasonable because the self-gravity is scale-free.  
Then the power index of the two-point correlation function is a good  
indicator representing the nonlinear dynamics of the self-gravity in this  
regime. And the power index is related to the clustering pattern of collisionless matters.  Hence it is very important to study what physical processes  
determine the power-index of the two-point correlation in the strongly nonlinear regime.  
Moreover it is interesting to analyze whether the power index in this regime  
depends on the initial conditions or not.    
  
These problems have been   
usually analyzed by   $N$-body simulations(Frenk, White, \& Davis 1983; Davis et.al. 1985; Suto 1993 and references therein).  
The method by $N$-body simulations is straightforward for tracking the  
evolution of the density fluctuations in the nonlinear regime. In the simulations, however, high spatial resolution is necessary for estimating   
the correlation functions and the mean relative peculiar velocity on very small scales in the strongly nonlinear regime with good accuracy(Jain 1995).   
But at the present time, the computer ability has not enough high resolution.  
Then we believe that the dynamics in the strongly nonlinear regime  
have never been completely verified by using the $N$-body simulations until   
now.  
  
There are other methods for the analysis of the nonlinear density fluctuations.  
One of them is the analysis by the BBGKY equations. The work by Davis \&  
Peebles (1977; hereafter DP) is a pioneer for the analysis by the BBGKY equations.  They showed the existence of the self-similar solutions for correlation functions under some assumptions.  Then it is shown that the power index $\gamma$ of the two-point spatial correlation function in the strongly nonlinear regime is related to the initial power index $n$ of the initial power spectrum $P(k)$ as follows:  
  
\begin{equation}  
\xi(r)\propto   
               r^{-\gamma}  (\xi\gg 1~:~\gamma=  
               \frac{\displaystyle 3(3+n)}{\displaystyle 5+n})  
\end{equation}  
  
This result is very interesting, but it also seems strange because  
in general systems might forget the initial memories in the strong nonlinear regime due to the nonlinearity of the self-gravity. One of the assumptions that DP adopted is called the stable condition.  
This condition is that the mean relative physical velocity in the strongly nonlinear regime is equal to zero. This condition was tested by $N$-body simulations  
(Efstathiou et al. 1988; Jain 1995) and this condition is not completely verified.  Furthermore the stability of the self-invariant solution in the strongly nonlinear regime, which DP derived, was investigated by the perturbation theory(Ruamsuwan \& Fry 1992) and it is found that the solution  
 is marginally stable.  
  
As for the physical process   
determining the power index in the strongly nonlinear regime, there are other analysis besides one proposed by DP.  One of them is given by Saslaw(1980).  He concluded that the power index $\gamma$ approaches to 2 by using the cosmic energy equation under some assumptions while some numerical simulations do not support this result(Frenk, White \& Davis 1983; Davis et al. 1985; Fry \& Melott 1985).    
  
There is another idea as follows; when the initial power spectrum has the sharp cut-off or the initial power spectrum is scale-free with negative and small initial power index, then there appear anywhere caustics of the density fields.  In these cases, the power index is irrespective of the detailed initial conditions after the first appearance of caustics on the small scales around the typical size of the thickness of caustics(in two-dimensional systems, they correspond to the filament structures of highly clustered matters).  The power index is determined by the type of the singularity which is classified in accord with the catastrophe theory.  This idea is verified in the one-dimensional system(Kotok \& Shandarin 1988; Gouda \& Nakamura 1988, 1989), the spherically symmetric systems(Gouda 1989) and the two-dimensional systems(Gouda 1996).  In these cases, it is suggested that $\gamma \approx 0$ on the small scales.    
  
As we can see from the above arguments, we believe that there are uncertainties about the physical processes which determine the value of the power index.  In this paper, we examine the conditions which determine the power index by analysing the scale-invariant solutions of the cosmic BBGKY equations.  The analysis of the nonlinear clustering in the strongly nonlinear regime by the BBGKY equations   
has advantage rather than that by the $N$-body simulations.  This is because the BBGKY equations directly deal with the statistical quantities such as correlation functions and this analysis is free from the artificial collisionality due to finite numbers of particles which might appear on the small scales in the $N$-body simulations.  However there is technically difficulty in solving the BBGKY equations; in our analysis, the BBGKY equations are translated to the moment equations by integrating the equations over velocity for convenience.  
Then these moment equations  for time evolution of the $N$-point spatial correlation function includes the terms including $N+1$-point spatial correlation functions.  
Furthermore $N$-th moment equation includes the terms including $N+1$-th   
moment. In general, these equations have infinite hierarchy and can not be closed at the lower order spatial correlation and moment.  Hence we should close the equations in taking some assumptions.  For example, DP took the following assumptions; the three-point spatial correlation function $\zeta$ is represented by the products of the two-point spatial correlation functions $\xi$ as follows:\begin{eqnarray}  
\zeta_{123} &=& Q(\xi_{12}\xi_{23}+\xi_{23}\xi_{31}+\xi_{31}\xi_{12}),  
\\   
\xi_{ik} &\equiv& \xi(x_i, x_k),~~~\zeta_{ijk} \equiv \zeta(x_i, x_j, x_k), 
\nonumber  
\end{eqnarray}  
where $Q$ is a constant.  Some observations suggest that this relation with $Q   
\sim 1$ holds at $\xi \sim 1$.  Furthermore DP assumed that the skewness of the velocity fields vanishes.  Adding one more assumption, DP closed the BBGKY equations.  As mentioned before, DP used the stable condition in deriving the power index $\gamma$ given by eq.(1). In this paper, we reexamine the scale-invariant solutions of the BBGKY equations and estimate the value of the power index and its dependence of the initial power index $n$ when the above assumptions and stable condition are changed.  And we analyze whether the stable condition is satisfied or not and how the mean relative peculiar velocity would be in real.  
Indeed, Jain(1995) claimed that the stable condition has not yet been verified by $N$-body simulations.  
  
Furthermore we investigate whether there is possibility that the power index in the strongly nonlinear regime does not depend on the initial power index $n$ even if the self-similarity of the solutions is satisfied.  Recently Padmanabhan(1995) suggested the possibility that the power index is independent of $n$ by using the pair conservation equations.    
  
In this paper, we examine the above problems by using the cosmological BBGKY equation in the strongly nonlinear regime.    
  
In $\S 2$, the cosmic BBGKY equations are briefly reviewed according to DP and Ruamsuwan \& Fry(1992).  We show the scale-invariant(power-law) solutions in the strongly nonlinear regime and their properties in $\S 3$.  Finally, we devote $\S 4$  
to conclusions and discussions.

\section{BASIC EQUATIONS}  
  
In this section, we briefly review the derivation of the cosmological BBGKY equations according to DP and Ruamsuwan \& Fry(1992).

\subsection{Cosmological BBGKY Equations}  
  Here we derive the BBGKY equations from the ensemble mean of the Vlasov equation in the expanding homogeneous and isotropic background universe.  
In this paper, we consider the only Einstein-de Sitter universe because  
we are interested in the scale-invariant solutions of the correlation   
functions and the self-similarity of the solutions and so it is necessary that the background universe has scale-free.  
  The $N$-body correlation  function is the statistical quantity which is   
given by the ensemble mean of the $N$-products of the one-body distribution functions. Then the BBGKY equations can be derived by the ensemble mean of the Vlasov  
equation.  
  The Vlasov equation for the one-body   
distribution function $f(\mbox{\boldmath$x,p$})$ is given by     
\begin{eqnarray}  
&& \frac{\partial f}{\partial t}  
+ \frac{p^{\alpha}}{ma^2} \frac{\partial f}{\partial x^\alpha}  
- m\frac{\partial \phi}{\partial x^\alpha}  
\frac{\partial f}{\partial p^\alpha} = 0, \\  
&& \phi(\mbox{\boldmath$x_1$})=\frac{Gm}{a}\int  
\frac{f(\mbox{\boldmath$x_2,p_2$})}{|\mbox{\boldmath$x_2-x_1$}|}  
d^3 x_2 d^3 p_2,  
\end{eqnarray}  
where $m$ is the mass of a particle, $a$ is the scale factor and  
$G$ is the gravitational constant.  
The ensemble mean of the Vlasov equation is  
\begin{equation}  
\frac{\partial \langle f \rangle}{\partial t}  
+ \frac{p^{\alpha}_1}{ma^2} \frac{\partial\langle f \rangle}  
{\partial x^\alpha_1}  
- m\langle\frac{\partial \phi}{\partial x^\alpha_1}  
\frac{\partial f}{\partial p^\alpha_1}\rangle = 0.  
\end{equation}  
  
We define the statistical functions which are given by the ensemble mean as follows:  
\begin{eqnarray}  
b(1)&=&\langle f(\mbox{\boldmath$x_1,p_1$})\rangle, \\  
\rho_2(1,2)&=&\langle f(\mbox{\boldmath$x_1,p_1$})  
 f(\mbox{\boldmath$x_2,p_2$})\rangle,\\  
\rho_3(1,2,3)&=&\langle f(\mbox{\boldmath$x_1,p_1$})  
 f(\mbox{\boldmath$x_2,p_2$}) f(\mbox{\boldmath$x_3,p_3$})\rangle, \\  
&\vdots& \nonumber  
\end{eqnarray}  
where $b(1)$ is a function of only the momentum because of the homogeneity in the background universe.  
  The ensemble mean of the Vlasov equation (5) is rewritten by using   
above functions(eqs.[6] $\sim$ [7]) as follows:  
\begin{equation}  
\frac{\partial b}{\partial t}  
+\frac{Gm^2}{a}\frac{\partial}{\partial p_1^\alpha}\int  
\frac{x_2^\alpha-x_1^\alpha}{|x_2-x_1|^3}\rho_2(1,2)d^3 x_2 d^3 p_2=0  
\end{equation}  
  This is the first BBGKY equation.   
As we can see from eq.(9), the time evolution of the one-body distribution function depends  
on the two-body correlation function.  
    
The following is the second BBGKY equation for the two-body correlation function:  
\begin{eqnarray}  
\frac{\partial \rho_2(1,2)}{\partial t}  
&=&\frac{\partial}{\partial t}\langle f(\mbox{\boldmath$x_1,p_1$})  
 f(\mbox{\boldmath$x_2,p_2$})\rangle \nonumber \\  
&=&-\frac{p_1^\alpha}{ma^2}\frac{\partial\rho_2(1,2)}{\partial x_1^\alpha}  
-\frac{Gm^2}{a}\frac{\partial}{\partial p_1^\alpha}\int  
\frac{x_{31}^\alpha}{x_{31}^3}\rho_3(1,2,3)d^3 x_3 d^3 p_3  
+(1\leftrightarrow 2),  
\end{eqnarray}  
\begin{equation}  
\frac{x_{31}^\alpha}{x_{31}^3}\equiv\frac{x_3^\alpha-x_1^\alpha}  
{|\mbox{\boldmath$x_3-x_1$}|^3}.  
\end{equation}  
  
  We can get the $N$-body correlation function by the same way as follows:   
\begin{eqnarray}  
\frac{\partial \rho_3(1,2,3)}{\partial t}  
&=&\frac{\partial}{\partial t}\langle f(\mbox{\boldmath$x_1,p_1$})  
f(\mbox{\boldmath$x_2,p_2$}) f(\mbox{\boldmath$x_3,p_3$})\rangle, \\  
\frac{\partial \rho_4(1,2,3,4)}{\partial t}  
&=&\frac{\partial}{\partial t}\langle f(\mbox{\boldmath$x_1,p_1$})  
f(\mbox{\boldmath$x_2,p_2$}) f(\mbox{\boldmath$x_3,p_3$})  
f(\mbox{\boldmath$x_4,p_4$})\rangle, \\  
&\vdots& \nonumber  
\end{eqnarray}  
In our analysis, however, we use the only second BBGKY equation as shown later.  
  
The second BBGKY equation is the time evolution equation of the   
two-body correlation function.  
  In this equation, the three-body correlation function is involved.  
  In general, the time evolution of the $N$-body correlation function depends on   
the $N+1$-body correlation function.  
  
  We define the following irreducible correlation functions, $c$ and $d$:  
\begin{eqnarray}  
\rho_2 (1,2)&\equiv&b(1)b(2)+c(1,2), \\  
\rho_3 (1,2,3)&\equiv&b(1)b(2)b(3)+b(1)c(2,3)+b(2)c(3,1)+b(3)c(1,2)+d(1,2,3).  
\end{eqnarray}  
Here $b, c$ and $d$ mean   
\begin{eqnarray}  
b(i)&=&b(\mbox{\boldmath$p_i$}),\\  
c(i,j)&=&c(\mbox{\boldmath$x_i,p_i,x_j,p_j$}),\\  
d(i,j,k)&=&d(\mbox{\boldmath$x_i,p_i,x_j,p_j,x_k,p_k$}).  
\end{eqnarray}  
  
The first and the second BBGKY equations are rewritten   
by using the above functions, respectively,  as follows:  
\begin{equation}  
\frac{\partial b(1)}{\partial t}  
+\frac{Gm^2}{a}\frac{\partial}{\partial p_1^\alpha}\int  
\frac{x_{21}^\alpha}{x_{21}^3}c(1,2)d^3 x_2 d^3 p_2=0  
~~~~~   ({\rm first~ BBGKY})   
\end{equation}  
\begin{eqnarray}  
\frac{\partial c(1,2)}{\partial t}  
&+&\frac{p_1^\alpha}{ma^2}\frac{\partial c(1,2)}{\partial x_1^\alpha}  
+\frac{Gm^2}{a}\frac{\partial b(1)}{\partial p_1^\alpha}\int  
\frac{x_{31}^\alpha}{x_{31}^3}c(2,3)d^3 x_3 d^3 p_3 \nonumber \\  
&+&\frac{Gm^2}{a}\frac{\partial }{\partial p_1^\alpha}\int  
\frac{x_{31}^\alpha}{x_{31}^3}d(1,2,3)d^3 x_3 d^3 p_3  
+(1\leftrightarrow 2)=0~~~   ({\rm second~ BBGKY})  
\end{eqnarray}

\subsection{Velocity Moment}  
  
We are interested in the power index of the two-point spatial correlation   
function in the strongly non-linear regime.  
Hence the equation which we use in our analysis is the second BBGKY equation.  
Moreover we use the velocity moment equations which are given by multiplying the second BBGKY equation by a  power of moment and integrate them over all moment  
arguments.  
The zeroth moment equation is given by    
\begin{equation}  
\frac{\partial}{\partial t}\int c(1,2)d^3 p_1 d^3 p_2  
+\frac{\partial}{\partial x^\alpha}\int \frac{1}{ma^2}p_{21}^\alpha  
 c(1,2)d^3 p_1 d^3 p_2 =0,  
~~~({\rm 0th~ moment})  
\end{equation}  
\begin{equation}  
x^\alpha \equiv x_{21}^\alpha \equiv x_2^\alpha - x_1^\alpha,~~~~~~~~  
p_{21}^\alpha \equiv p_2^\alpha - p_1^\alpha.  
\end{equation}  
  
As we can see, the time evolution of the zeroth moment  depends on   
the first moment in the BBGKY equation.   
   
The first moment equation is given by   
\begin{eqnarray}  
\frac{\partial}{\partial t}\int c(1,2)p_{21}^\beta d^3 p_1 d^3 p_2  
&+&\frac{\partial}{\partial x^\alpha}\int\frac{1}{ma^2}  
p_{21}^\alpha p_{21}^\beta c(1,2)d^3 p_1 d^3 p_2 \nonumber \\  
&+& \int \frac{Gm^2}{a}\{b(1)c(2,3)+d\}\frac{x_{31}^\beta}{x_{31}^3}  
d^3 x_3 d^3 p_3 d^3 p_1 d^3 p_2 \nonumber \\  
&-& \int \frac{Gm^2}{a}\{b(2)c(3,1)+d\}\frac{x_{32}^\beta}{x_{32}^3}  
d^3 x_3 d^3 p_3 d^3 p_1 d^3 p_2 =0. \nonumber \\  
&&~~~~~~~~~~~~~~~~~  
~~~~~~~~~~~~~~~({\rm 1st~ moment})  
\end{eqnarray}  
  
This equation includes the second moment and then we need the second moment equation as follows:  
\begin{eqnarray}  
\frac{\partial}{\partial t}\int c(1,2)p_{21}^\beta p_{21}^\gamma   
d^3 p_1 d^3 p_2  
&+&\frac{\partial}{\partial x^\alpha}\int\frac{1}{ma^2}  
p_{21}^\alpha p_{21}^\beta p_{21}^\gamma c(1,2)d^3 p_1 d^3 p_2 \nonumber \\  
&+& \int \frac{Gm^2}{a}\{b(1)c(2,3)+d\}  
\{p_{21}^\gamma\frac{x_{31}^\beta}{x_{31}^3}  
-p_{21}^\beta\frac{x_{31}^\gamma}{x_{31}^3}\}  
d^3 x_3 d^3 p_3 d^3 p_1 d^3 p_2 \nonumber \\  
&-& \int \frac{Gm^2}{a}\{b(2)c(3,1)+d\}  
\{p_{21}^\gamma\frac{x_{32}^\beta}{x_{32}^3}  
-p_{21}^\beta\frac{x_{32}^\gamma}{x_{32}^3}\}  
d^3 x_3 d^3 p_3 d^3 p_1 d^3 p_2 =0.  \nonumber \\  
&&~~~~~~~~~~~~~~~~~~~~~~~~~~  
~~~~~~~~~~~~~~~~~~~~~~~~({\rm 2nd~ moment})  
\end{eqnarray}  
  
As we can see above, the time evolution of the $N$-th moment depends on the   
$N+1$-th moment. So we should take assumptions in order to close these   
equations. DP used the assumption that the skewness of the velocity   
is equal to 0.  
  In this paper, we do not assume about the skewness of the velocity fields in order to study   
the relation between the skewness and the two-point correlation function.  
  
  Here we define the two-point correlation function $\xi$, the mean relative peculiar velocity $\langle v^{\alpha} \rangle$, the relative peculiar velocity dispersion $\Pi, \Sigma$ and the mean third moment $\langle v^\alpha v^\beta  
v^\gamma \rangle$ as follows:  
\begin{eqnarray}  
\bar{n}^2 a^6 \xi  \label{def1} &\equiv&\int c(2,1) d^3 p_1 d^3 p_2,  \\  
\bar{n}^2 a^6 (1+\xi)ma \langle v^\alpha \rangle    
&\equiv& \int c(2,1) p^{\alpha}_{21} d^3 p_1 d^3 p_2, \label{def2} \\  
\bar{n}^2 a^6 (1+\xi)(ma)^2   
[\Pi P_\parallel^{\alpha \beta} + \Sigma P_\perp^{\alpha \beta}]  
&\equiv&   
\int c(2,1)p_{21}^\alpha p_{21}^\beta d^3 p_1 d^3 p_2, \\  
\langle v^\alpha v^\beta v^\gamma \rangle  
&\equiv& \frac{\displaystyle  
              \frac{\displaystyle 1}{\displaystyle (ma)^3}  
              \int \rho_2 p_{21}^\alpha p_{21}^\beta p_{21}^\gamma  
              d^3 p_1 d^3 p_2  
             }   
              {\displaystyle \int \rho_2 d^3 p_1 d^3 p_2} \nonumber\\  
&=& \frac { \displaystyle   
         \int c(2,1) p_{21}^\alpha p_{21}^\beta p_{21}^\gamma  
         d^3 p_1 d^3 p_2  
        }  
         {\displaystyle (ma)^3 \bar{n}^2 a^6 (1+\xi)},   
\label{def4}  
\end{eqnarray}  
where  
\begin{equation}  
 P_\parallel^{\alpha \beta}=\frac{x^\alpha x^\beta}{x^2},~~~~~~~  
 P_\perp^{\alpha \beta}=\delta^{\alpha \beta}-\frac{x^\alpha x^\beta}{x^2}.  
\end{equation}  
Here $\bar{n}$ is the mean density of the universe.  
The $\xi$ which is defined in eq.(25) is equivalent to  
the two-point correlation function $\xi_2(x)$ which is  defined by another familiar definition as follows:   
\begin{eqnarray}	  
\xi_2(x)&\equiv&\langle\delta(\mbox{\boldmath$r$})  
        \delta(\mbox{\boldmath$r+x$})\rangle_{\mbox{\boldmath$r,|x|$}=x}  
                                          \nonumber \\  
&=&\frac{1}{\langle \rho \rangle^2}  
    \left\langle (\rho(\mbox{\boldmath$r$})-\langle \rho \rangle )  
                 (\rho(\mbox{\boldmath$r+x$})-\langle \rho \rangle )  
    \right\rangle_{\mbox{\boldmath$r,|x|$}=x}  
                                          \nonumber \\  
&=&\frac{1}{\bar{n}^2 a^6}  
    \left[  
     \langle \rho(\mbox{\boldmath$r$})\rho(\mbox{\boldmath$r+x$})\rangle  
  -  \langle \rho \rangle \langle \rho \rangle   
    \right]  
                                          \nonumber \\  
&=&\frac{1}{\bar{n}^2 a^6}\int(\rho_2(1,2)-b(1)b(2))d^3 p_1 d^3 p_2  
                                          \nonumber \\  
&=&\frac{1}{\bar{n}^2 a^6}\int c(1,2) d^3 p_1 d^3 p_2  
                                          \nonumber \\  
&=& \xi(x_{21}).  
\end{eqnarray}  
where $\langle \rho \rangle$ is the mean density of the universe.  
  
In eq.(26), $\langle v^{\alpha} \rangle$ is the mean relative peculiar velocity  
given by   
\begin{eqnarray}  
   \langle v^\alpha \rangle   
&\equiv& \frac{ \displaystyle  
                \int \rho_2(2,1)  
                \frac{\displaystyle 1}{\displaystyle ma}  
                p_{21}^\alpha d^3 p_1 d^3 p_2   
               }  
                {\displaystyle \int \rho_2(2,1) d^3 p_1 d^3 p_2 }  
                                          \nonumber \\  
&=& \frac{ \displaystyle  
           \int c(2,1)p_{21}^\alpha d^3 p_1 d^3 p_2  
          }  
           {\displaystyle \bar{n}^2 a^6 (ma)(1+\xi)}.  
\end{eqnarray}  
  
In eq.(27), $\Pi$ and $\Sigma$ are the parallel and transverse correlated parts of the relative peculiar velocity dispersion of correlated particles, respectively.  
   
Here we define the skewness as follows:  
  
\begin{equation}  
 s^{\alpha \beta \gamma}  
\equiv  
\langle(v-\langle v \rangle)^\alpha \rangle  
\langle(v-\langle v \rangle)^\beta \rangle  
\langle(v-\langle v \rangle)^\gamma \rangle.  
\end{equation}  
  
{}From the symmetry of the universe(homogeneity and isotropy), we can write   
$\langle v^\alpha \rangle = \langle v \rangle {x^\alpha}/{x}$.  
Hence,  
\begin{equation}  
\langle v^\alpha v^\beta v^\gamma \rangle  
=-2\langle v \rangle^3 \frac{x^\alpha x^\beta x^\gamma}{x^3}  
+\langle v \rangle  
\left\{\frac{x^\alpha}{x}\langle v^\beta v^\gamma \rangle  
+\frac{x^\beta}{x}\langle v^\gamma v^\alpha \rangle  
+\frac{x^\gamma}{x}\langle v^\alpha v^\beta \rangle  
\right\}+s^{\alpha \beta \gamma}.  
             \label{skew1}  
\end{equation}  
  
Furthermore the skewness can be written by the symmetry of   
the universe as follows:  
\begin{equation}  
s^{\alpha \beta \gamma}=  
s_\parallel ~ P_{ppp}^{\alpha \beta\gamma}  
 + s_\perp ~ P_{ptt}^{\alpha \beta \gamma},  
\end{equation}  
where the subscripts $p$ and $t$ represent the parallel and transverse component of the two particles, respectively.   
  
\begin{equation}  
P_{ppp}^{\alpha \beta\gamma}  
=\frac{x^\alpha x^\beta x^\gamma}{x^3},~~~~  
 P_{ptt}^{\alpha \beta \gamma}  
=  \frac{x^\alpha}{x}\delta^{\beta \gamma}  
  +\frac{x^\beta}{x}\delta^{\gamma \alpha}  
  +\frac{x^\gamma}{x}\delta^{\alpha \beta}   
  -3\frac{x^\alpha x^\beta x^\gamma}{x^3}.  
\end{equation}

$P_{ppt}$ and $ P_{ttt}$ vanish because of the symmetry of   
the universe.  
  
{}From eqs.(28) and (33), we can get the following equations:  
\begin{eqnarray}  
\int c(2,1)p_{21}^\alpha p_{21}^\beta p_{21}^\gamma d^3 p_1 d^3 p_2   
= \bar{n}^2 a^6 (1+\xi)(ma)^3  
\{   
[3\langle v \rangle(\Pi-\Sigma)-2\langle v \rangle^3]  
\frac{x^\alpha x^\beta x^\gamma}{x^3} \nonumber \\  
+ \langle v \rangle(\Sigma +\frac{2\langle v_1^2 \rangle}{3(1+\xi)})  
(  \frac{x^\alpha}{x}\delta^{\beta \gamma}  
  +\frac{x^\beta}{x}\delta^{\gamma \alpha}  
  +\frac{x^\gamma}{x}\delta^{\alpha \beta} )  
+s^{\alpha \beta \gamma.}  
\}  
\label{skew2}  
\end{eqnarray}  
  
Then moment equations are following:  
  
\begin{equation}  
\bar{n}^2 a^6 \frac{\partial \xi}{\partial t}  
+\frac{\bar{n}^2 a^6 (ma)}{ma^2}\frac{\partial}{\partial x^\alpha}  
[(1+\xi)\langle v^\alpha \rangle]=0,  
~~~~~~~({\rm 0th~moment})  
\end{equation}  
\begin{eqnarray}  
\bar{n}^2 a^6 \frac{\partial}{\partial t}[(1+\xi)ma\langle v^\beta \rangle]  
&+&\frac{\bar{n}^2 a^6 (ma)^2}{ma^2}\frac{\partial}{\partial x^\alpha}  
\left[  
(1+\xi)[\Pi P_\parallel^{\alpha \beta} + \Sigma P_\perp^{\alpha \beta}]  
\right] \nonumber \\  
&+&\frac{Gm^2}{a}\bar{n}a^3 \int c(2,3)\frac{x_{31}^\beta}{x_{31}^3}  
d^3 x_3 d^3 p_3 d^3 p_2 \nonumber \\  
&-&\frac{Gm^2}{a}\bar{n}a^3 \int c(3,1)\frac{x_{32}^\beta}{x_{32}^3}  
d^3 x_3 d^3 p_3 d^3 p_1 \nonumber \\  
&+&\frac{Gm^2}{a}\bar{n}^3 a^9  
 \int \zeta (\frac{x_{31}^\beta}{x_{31}^3}-\frac{x_{32}^\beta}{x_{32}^3})  
d^3 x_3 =0,~~~~({\rm 1st~moment})  
                                  \label{firstm}  
\end{eqnarray}  
\begin{eqnarray}  
\lefteqn{  
         \bar{n}^2 a^6 \frac{\partial}{\partial t}  
         \left[  
               (ma)^2 (1+\xi)  
               [  \Pi P_\parallel^{\alpha \beta}   
                + \Sigma P_\perp^{\alpha \beta}  ]   
         \right]  
        } \nonumber \\  
&&~~ + \frac{ \bar{n}^2 a^6 (ma)^3}{ma^2}\frac{\partial}{\partial x^\alpha}  
       (1+\xi) [3\langle v \rangle(\Pi-\Sigma)-2\langle v \rangle^3]  
       \frac{x^\alpha x^\beta x^\gamma}{x^3} \nonumber \\  
&&~~ + \frac{ \bar{n}^2 a^6 (ma)^3}{ma^2}\frac{\partial}{\partial x^\alpha}  
       (1+\xi)\langle v \rangle  
       (\Sigma +\frac{2\langle v_1^2 \rangle}{3(1+\xi)})  
       (  \frac{x^\alpha}{x}\delta^{\beta \gamma}  
         +\frac{x^\beta}{x}\delta^{\gamma\alpha}  
         +\frac{x^\gamma}{x}\delta^{\alpha \beta} )\nonumber \\  
&&~~ + \frac{ \bar{n}^2 a^6 (ma)^3}{ma^2}\frac{\partial}{\partial x^\alpha}  
       s^{\alpha \beta \gamma}\nonumber \\  
&&~~ + \frac{Gm^2}{a}\bar{n}a^3 \int c(2,3)  
       [  p_{21}^\gamma\frac{x_{31}^\beta}{x_{31}^3}  
         +p_{21}^\beta\frac{x_{31}^\gamma}{x_{31}^3}]  
       d^3 x_3 d^3 p_3 d^3 p_2 \nonumber \\  
&&~~ - \frac{Gm^2}{a}\bar{n}a^3 \int c(3,1)  
       [  p_{21}^\gamma\frac{x_{32}^\beta}{x_{32}^3}  
         +p_{21}^\beta\frac{x_{32}^\gamma}{x_{32}^3}]  
       d^3 x_3 d^3 p_3 d^3 p_1 \nonumber \\  
&&~~ + \frac{Gm^2}{a}\bar{n}a^3 \int d  
       [  p_{21}^\gamma   
          (\frac{x_{31}^\beta}{x_{31}^3}-\frac{x_{32}^\beta}{x_{32}^3})  
        + p_{21}^\beta (\frac{x_{31}^\gamma}{x_{31}^3}  
        - \frac{x_{32}^\gamma}{x_{32}^3})  
       ]  
       d^3 x_3 d^3 p_3 d^3 p_1 d^3 p_2 =0. \nonumber \\  
&&~~~~~~~~~~~~~~~~~~~~~~~~~~~~~~~~~~~  
~~~~~~~~~~~~~~~~~~~~~~~~~~~~~~~~~({\rm 2nd~moment})  
                    \label{secondm}  
\end{eqnarray}  
where $\zeta$ is the three-point correlation function defined by  
\begin{equation}  
\bar{n}^3 a^9 \zeta  
\equiv  
\int d d^3 p_1 d^3 p_2 d^3 p_3.   
\end{equation}  
  
\subsection{Contraction of the Equation}  
  
  The zeroth moment and first moment equations are the time evolution   
equations of the $\xi$ and $\langle v \rangle$, respectively.   
  The second moment equation is the time evolution equation of the   
$\Pi,\Sigma$.  
For convenience, we transform these equations by taking divergence of the first moment equation and by operating the following two operators to the second moment equation,   
\begin{equation}  
\frac{\partial}{\partial x^\beta}\frac{\partial}{\partial x^\gamma}~,~~~~~~~~  
\Delta^{\beta \gamma}=\frac{1}{2}(\delta^{\beta \gamma}  
-\frac{x^\beta x^\gamma}{x^2}).  
\end{equation}  
   
Hence we get two equations from the second moment equation.  We call them  
contraction 1and contraction 2 equations hereafter, shown later in eqs.(50) and (51) , respectively. Here we assume the following relation,  
\begin{equation}  
\zeta_{123}=Q(\xi_{12}\xi_{23}+\xi_{23}\xi_{31}+\xi_{31}\xi_{12}).  
\end{equation}  
  
Then the fifth term of the first moment equation (\ref{firstm}) is rewritten by  
\begin{eqnarray}  
\frac{Gm^2}{a}\bar{n}^3 a^9  
     \frac{\partial}{\partial x^\beta} \int \zeta   
     (\frac{x_{31}^\beta}{x_{31}^3}-\frac{x_{32}^\beta}{x_{32}^3}) d^3 x_3  
&=& 2  \frac{Gm^2}{a}\bar{n}^3 a^9  
       \frac{\partial}{\partial x^\beta} \int \zeta   
       (\frac{x_{31}^\beta}{x_{31}^3})d^3 x_3   
                          \nonumber \\  
&=&  2 \frac{Gm^2}{a}\bar{n}^3 a^9 Q  
       \frac{\partial}{\partial x^\beta} \int    
       \frac{x_{31}^\beta}{x_{31}^3}d^3 x_3   
       (\xi_{12}\xi_{23}+\xi_{23}\xi_{31}+\xi_{31}\xi_{12})  
                          \nonumber \\  
&=&  2 \frac{Gm^2}{a}\bar{n}^3 a^9 Q  
       \frac{\partial}{\partial x^\beta} \int    
       \frac{x_{31}^\beta}{x_{31}^3}d^3 x_3   
       (\xi_{12}+\xi_{31})\xi_{23}  
                          \nonumber \\  
&\equiv&  2 \frac{Gm^2}{a}\bar{n}^3 a^9 Q  
       \frac{\partial}{\partial x^\beta}  
       [x^\beta T]  
                          \nonumber \\  
&=&      2 \frac{Gm^2}{a}\bar{n}^3 a^9 Q  
       [ \frac{\partial T}{\partial x}x +T]  
                          \nonumber \\  
&=&      2 \frac{Gm^2}{a}\bar{n}^3 a^9 Q  
       \frac{1}{x^2}\frac{\partial}{\partial x} [ x^3 T]  
                          \nonumber \\  
&=&      2 \frac{Gm^2}{a}\bar{n}^3 a^9 Q M  
       \frac{1}{x^2}\frac{\partial}{\partial x} [ x^3 \xi^2],  
\end{eqnarray}  
where  
\begin{equation}  
          x^\beta T   
\equiv    x^\beta \xi(x)^2 M  
\equiv    \int   
          \frac{x_{31}^\beta}{x_{31}^3}  
          \{ \xi_{12} + \xi_{31} \} \xi_{23}   
          d^3 x_3,~~~~~~~~  x\equiv x_{21}.   
\end{equation}  
This definition is well defined for the vector component of  
$\mbox{\boldmath$x$}$    
and $\mbox{\boldmath$x$}_{31}$ 
because of the symmetry of the universe.  
  
The seventh term in eq.(39) is rewritten by  
\begin{eqnarray}  
\lefteqn{  
 \frac{Gm^2}{a}   
     \frac{\partial^2}{\partial x^\beta \partial x^\gamma}  
     \int d[p_{21}^\gamma   
     (\frac{x_{31}^\beta}{x_{31}^3}-\frac{x_{32}^\beta}{x_{32}^3})  
   +        p_{21}^\beta   
     (\frac{x_{31}^\gamma}{x_{31}^3}-\frac{x_{32}^\gamma}{x_{32}^3})]  
     d^3 x_3 d^3 p_3 d^3 p_1 d^3 p_2  
}   
                          \nonumber \\  
&=&  4 \frac{Gm^2}{a}   
     \frac{\partial^2}{\partial x^\beta \partial x^\gamma}  
     \int dp_{21}^\gamma   
     \frac{x_{31}^\beta}{x_{31}^3}  
     d^3 x_3 d^3 p_3 d^3 p_1 d^3 p_2   
                          \nonumber \\  
&=&  4 \frac{Gm^2}{a}(ma)\bar{n}^3 a^9 Q^*  
     \frac{\partial^2}{\partial x^\beta \partial x^\gamma}  
     \int   
     \frac{x_{31}^\beta}{x_{31}^3}  
     [  \langle v_{21}^\gamma \rangle \xi_{12}\xi_{23}  
    +   (\langle v_{23}^\gamma \rangle + \langle v_{31}^\gamma \rangle)  
        \xi_{23}\xi_{31}  
    +   \langle v_{21}^\gamma \rangle \xi_{31}\xi_{12} ]  
     d^3 x_3   
                          \nonumber \\  
&=&  4 \frac{Gm^2}{a}(ma)\bar{n}^3 a^9 Q^*  
     \frac{\partial^2}{\partial x^\beta \partial x^\gamma}  
     \int   
     \frac{x_{31}^\beta}{x_{31}^3}  
     [  x^\gamma  \frac{\langle v_{21} \rangle}{x} \xi_{12}\xi_{23}  
    +   \{(x^\gamma -z^\gamma)\frac{\langle v_{23} \rangle}{|x-z|}  
          + z^\gamma \frac{\langle v_{31} \rangle}{z} \}  
        \xi_{23}\xi_{31} ]  
     d^3 x_3   
                          \nonumber \\  
&=&  4 \frac{Gm^2}{a}(ma)\bar{n}^3 a^9 Q^*  
     \frac{\partial^2}{\partial x^\beta \partial x^\gamma}  
     \int   
     \frac{x_{31}^\beta}{x_{31}^3}  
       x^\gamma \{    \frac{\langle v_{21} \rangle}{x} \xi_{12}  
                   +  \frac{\langle v_{23} \rangle}{|x-z|}\xi_{31}   
                \} \xi_{23}  
                         d^3 x_3    \nonumber \\  
&&~~~~+      4 \frac{Gm^2}{a}(ma)\bar{n}^3 a^9 Q^*  
     \frac{\partial^2}{\partial x^\beta \partial x^\gamma}  
     \int   
     \frac{x_{31}^\beta}{x_{31}^3}  
     z^\gamma \{    \frac{\langle v_{31} \rangle}{z}   
                 -  \frac{\langle v_{23} \rangle}{|x-z|}  
               \} \xi_{23}\xi_{31}    
                             d^3 x_3   
\end{eqnarray}  
where  
\begin{equation}  
\frac{Q^*}{Q}  
 \bar{n}^3 a^9 \zeta ma \langle v_{21} \rangle  
\equiv   
\int d p_{21}^\alpha d^9 p   
\label{zetafirst}  
\end{equation}  
  
In general this relation (eq.[46]) is not satisfied.  
But DP showed the existence of $d$ which satisfies this relation.  
Furthermore, in the strongly nonlinear regime, which we are interested in, we find from the dimensional analysis that this relation is correct in general.   
  
  We can see from eqs.(39) and (45) that the seventh term of the second moment equation is transformed by operator $\Delta^{\beta \gamma}$ as follows,  
\begin{eqnarray}  
\lefteqn{  
 \frac{Gm^2}{a}   
     \Delta^{\beta \gamma}  
     \int d[p_{21}^\gamma   
     (\frac{x_{31}^\beta}{x_{31}^3}-\frac{x_{32}^\beta}{x_{32}^3})  
   +        p_{21}^\beta   
     (\frac{x_{31}^\gamma}{x_{31}^3}-\frac{x_{32}^\gamma}{x_{32}^3})]  
     d^3 x_3 d^3 p_3 d^3 p_1 d^3 p_2  
}   
                          \nonumber \\  
&=&  4 \frac{Gm^2}{a}   
     \Delta^{\beta \gamma}  
     \int dp_{21}^\gamma   
     \frac{x_{31}^\beta}{x_{31}^3}  
     d^3 x_3 d^3 p_3 d^3 p_1 d^3 p_2   
                          \nonumber \\  
&=&  4 \frac{Gm^2}{a}(ma)\bar{n}^3 a^9 Q^*  
     \Delta^{\beta \gamma}  
     \int   
     \frac{x_{31}^\beta}{x_{31}^3}  
       x^\gamma \{    \frac{\langle v_{21} \rangle}{x} \xi_{12}  
                   +  \frac{\langle v_{23} \rangle}{|x-z|}\xi_{31}   
                \} \xi_{23}  
                         d^3 x_3    \nonumber \\  
&&~~~~+      4 \frac{Gm^2}{a}(ma)\bar{n}^3 a^9 Q^*  
     \Delta^{\beta \gamma}  
     \int   
     \frac{x_{31}^\beta}{x_{31}^3}  
     z^\gamma \{    \frac{\langle v_{31} \rangle}{z}   
                 -  \frac{\langle v_{23} \rangle}{|x-z|}  
               \} \xi_{23}\xi_{31}    
                             d^3 x_3.   
\end{eqnarray}  
  
  Finally we obtain the following four equations:  
\begin{equation}  
\bar{n}^2 a^6 \frac{\partial \xi}{\partial t}  
+\frac{\bar{n}^2 a^6(ma)}{ma^2}\frac{1}{x^2}  
\frac{\partial}{\partial x}[x^2(1+\xi)\langle v \rangle]=0.  
~~~~~~~({\rm0th~moment})  
\end{equation}  
\begin{eqnarray}  
\bar{n}^2 a^6 \frac{\partial}{\partial t}\frac{1}{x^2}  
\frac{\partial}{\partial x}[x^2(1+\xi)(ma)\langle v \rangle]  
&+&\frac{\bar{n}^2 a^6(ma)^2}{ma^2}\frac{1}{x^2}\frac{\partial}{\partial x}  
\{\frac{\partial}{\partial x}(x^2(1+\xi) \Pi) - 2x(1+\xi) \Sigma \}  
                               \nonumber \\  
&+&\frac{Gm^2}{a}\bar{n}^3 a^9 8\pi \xi  
                               \nonumber \\  
&+&      2 \frac{Gm^2}{a}\bar{n}^3 a^9 Q M  
         \frac{1}{x^2}\frac{\partial}{\partial x} [ x^3 \xi^2]=0  
                                \nonumber \\  
&&~~~~~~~~~~~~~~~~~~~  
~~~~~~~~~~~~~~~~~~({\rm 1st~moment})  
\end{eqnarray}  
\begin{eqnarray}  
\lefteqn{  
          \bar{n}^2 a^6 \frac{\partial}{\partial t}  
          \frac{(ma)^2}{x^2}\frac{\partial}{\partial x}  
          \left[  
                 \frac{\partial}{\partial x}  
                 [x^2(1+\xi)\Pi]-2x(1+\xi) \Sigma   
          \right]  
         }     \nonumber \\  
&&~~~~+~  \frac{ \bar{n}^2 a^6 (ma)^3}{ma^2}\frac{1}{x^2}  
       \frac{\partial^3}{\partial x^3}  
        \left[  
        x^2(1+\xi) \{3\langle v \rangle(\Pi-\Sigma)-2\langle v \rangle^3\}  
        \right]    
     \nonumber \\  
&&~~~~+~  \frac{ \bar{n}^2 a^6 (ma)^3}{ma^2}  
       \frac{3}{x^2}\frac{\partial}{\partial x}  
       \frac{1}{x}\frac{\partial}{\partial x}  
       \left[  
             x^4\frac{\partial}{\partial x}\frac{1}{x}  
             (1+\xi)\langle v \rangle  
             \{\Sigma   +        \frac{2\langle v_1^2 \rangle}{3(1+\xi)} \}  
       \right]  
                                  \nonumber \\  
&&~~~~+~  \frac{ \bar{n}^2 a^6 (ma)^3}{ma^2}\frac{1}{x^2}  
       \frac{\partial^3}{\partial x^3}  
       [x^2 (1+\xi)s_\parallel]   
                                   \nonumber \\  
&&~~~~+~  \frac{16\pi G m^2}{a}  
       \bar{n} a^3 \bar{n}^2 a^6 \frac{1}{x^2}\frac{\partial}{\partial x}  
          [x^2 (1+\xi)(ma)\langle v \rangle]  
                                   \nonumber \\  
&&~~~~+~  4 \frac{Gm^2}{a}(ma)\bar{n}^3 a^9 Q^*  
       \frac{\partial^2}{\partial x^\beta \partial x^\gamma}  
       \int   
       \frac{x_{31}^\beta}{x_{31}^3}  
       x^\gamma \{ \frac{\langle v_{21} \rangle}{x} \xi_{12}  
    +  \frac{\langle v_{23} \rangle}{|x-z|}\xi_{31} \} \xi_{23}  
       d^3 x_3   
                          \nonumber \\  
&&~~~~+~  4 \frac{Gm^2}{a}(ma)\bar{n}^3 a^9 Q^*  
       \frac{\partial^2}{\partial x^\beta \partial x^\gamma}  
       \int   
       \frac{x_{31}^\beta}{x_{31}^3}  
       z^\gamma \{  \frac{\langle v_{31} \rangle}{z}   
      -\frac{\langle v_{23} \rangle}{|x-z|} \} \xi_{23}\xi_{31}    
       d^3 x_3=0   
                          \nonumber \\  
&&~~~  
~~~~~~~~~~~~~~~~~~~({\rm 2nd~moment:contraction~ 1})  
                                      \label{shuku3}  
\end{eqnarray}  
\begin{eqnarray}  
\bar{n}^2 a^6 \frac{\partial}{\partial t}  
[(ma)^2(1+\xi) \Sigma]   
&+& \frac{ \bar{n}^2 a^6 (ma)^3}{ma^2}\frac{1}{x^4}  
\frac{\partial}{\partial x}  
\left[  
x^4(1+\xi) \langle v \rangle   
\{\Sigma + \frac{2\langle v_1^2 \rangle}{3(1+\xi)}\}  
\right]       \nonumber \\  
&+&\frac{ \bar{n}^2 a^6 (ma)^3}{ma^2}\frac{1}{x^2}  
\frac{1}{x^4}  
\frac{\partial}{\partial x}  
[x^4 (1+\xi)s_\perp]  \nonumber \\  
&+&\frac{Gm^2}{a}\bar{n}a^3   
          \Delta^{\beta \gamma}  
          \int c(2,3)  
          [   p_{21}^\gamma\frac{x_{31}^\beta}{x_{31}^3}  
           +  p_{21}^\beta\frac{x_{31}^\gamma}{x_{31}^3}]  
          d^3 x_3 d^3 p_3 d^3 p_2                     \nonumber \\  
&-&   \frac{Gm^2}{a}\bar{n}a^3   
          \Delta^{\beta \gamma}  
          \int c(3,1)  
          [   p_{21}^\gamma\frac{x_{32}^\beta}{x_{32}^3}  
           +  p_{21}^\beta\frac{x_{32}^\gamma}{x_{32}^3}]  
          d^3 x_3 d^3 p_3 d^3 p_1                      \nonumber \\  
&+&  4 \frac{Gm^2}{a}(ma)\bar{n}^3 a^9 Q^*  
     \Delta^{\beta \gamma}  
     \int   
     \frac{x_{31}^\beta}{x_{31}^3}  
       x^\gamma \{ \frac{\langle v_{21} \rangle}{x} \xi_{12}  
    +  \frac{\langle v_{23} \rangle}{|x-z|}\xi_{31} \} \xi_{23}  
     d^3 x_3   
                          \nonumber \\  
&+&  4 \frac{Gm^2}{a}(ma)\bar{n}^3 a^9 Q^*  
     \Delta^{\beta \gamma}  
     \int   
     \frac{x_{31}^\beta}{x_{31}^3}  
     z^\gamma \{  \frac{\langle v_{31} \rangle}{z}   
    -\frac{\langle v_{23} \rangle}{|x-z|} \} \xi_{23}\xi_{31}    
     d^3 x_3 =0.  
                          \nonumber \\  
&&~~~~~~~~~~~~~({\rm 2nd~moment: contraction~ 2})  
                                     \label{shuku4}  
\end{eqnarray}

We are interested in the strongly nonlinear regime and then the solutions of the above equations are expected to obey the power law because the self-gravity is scale-free.  We assume that  $\xi$ is represented by the power law form as follows:  
\begin{equation}  
\xi = \xi_0 a^\beta x^{-\gamma}  
\end{equation}  
where $\xi_0, \beta$, and $\gamma$ are constants.  
  
  Then we obtain from the dimensional analysis in eq.(48) in the strongly nonlinear regime.   
\begin{equation}  
\langle v \rangle = -h\dot{a}x  
\end{equation}  
where $h$ is a constant.  
  
  Then the sixth term of the second moment equation(contraction 1, eq.50),  
is rewritten by  
\begin{eqnarray}  
   \lefteqn{4 \frac{Gm^2}{a}(ma)\bar{n}^3 a^9 Q^*  
             \frac{\partial^2}{\partial x^\beta \partial x^\gamma}  
             \int   
             \frac{x_{31}^\beta}{x_{31}^3}  
             [  x^\gamma \{ \frac{\langle v_{21} \rangle}{x} \xi_{12}  
          +  \frac{\langle v_{23} \rangle}{|x-z|}\xi_{31} \} \xi_{23} ]  
              d^3 x_3  
}  
                          \nonumber \\  
&=&  4 \frac{Gm^2}{a}(ma)\bar{n}^3 a^9 Q^* (-\dot{a}h)  
     \frac{\partial^2}{\partial x^\beta \partial x^\gamma}  
     \int   
     \frac{x_{31}^\beta}{x_{31}^3}  
       x^\gamma \{ \xi_{12} + \xi_{31} \} \xi_{23}   
     d^3 x_3   
                          \nonumber \\  
&=&  4 \frac{Gm^2}{a}(ma)\bar{n}^3 a^9 Q^* (-\dot{a}h)  
     \frac{\partial^2}{\partial x^\beta \partial x^\gamma}  
     [x^\gamma x^\beta T]  
                          \nonumber \\  
&=&  4 \frac{Gm^2}{a}(ma)\bar{n}^3 a^9 Q^* (-\dot{a}h)  
     \frac{1}{x^2}  \frac{\partial^2}{\partial x^2}  
     [x^4 T]  
                          \nonumber \\  
&=&  4 \frac{Gm^2}{a}(ma)\bar{n}^3 a^9 Q^* M (-\dot{a}h)  
     \frac{1}{x^2}  \frac{\partial^2}{\partial x^2}  
     [x^4 \xi^2].  
\end{eqnarray}  
  
  The seventh term of the second moment equation(contraction 1, eq.50)  
is rewritten by   
\begin{equation}  
     4 \frac{Gm^2}{a}(ma)\bar{n}^3 a^9 Q^*  
     \frac{\partial^2}{\partial x^\beta \partial x^\gamma}  
     \int   
     \frac{x_{31}^\beta}{x_{31}^3}  
     z^\gamma \{    \frac{\langle v_{31} \rangle}{z}   
                 -  \frac{\langle v_{23} \rangle}{|x-z|}  
               \} \xi_{23}\xi_{31}    
                             d^3 x_3=0.   
\end{equation}  
  
The sixth term of the second moment equation(contraction 2, eq.51) is rewritten by ,  
\begin{eqnarray}  
\lefteqn{  
         4 \frac{Gm^2}{a}(ma)\bar{n}^3 a^9 Q^*  
             \Delta^{\beta \gamma}  
             \int   
             \frac{x_{31}^\beta}{x_{31}^3}  
             [  x^\gamma \{ \frac{\langle v_{21} \rangle}{x} \xi_{12}  
          +  \frac{\langle v_{23} \rangle}{|x-z|}\xi_{31} \} \xi_{23} ]  
              d^3 x_3.  
       }   
                          \nonumber \\  
&=&  4 \frac{Gm^2}{a}(ma)\bar{n}^3 a^9 Q^* (-\dot{a}h)  
             \Delta^{\beta \gamma}  
     \int   
     \frac{x_{31}^\beta}{x_{31}^3}  
       x^\gamma \{ \xi_{12} + \xi_{31} \} \xi_{23}   
     d^3 x_3   
                          \nonumber \\  
&=&  4 \frac{Gm^2}{a}(ma)\bar{n}^3 a^9 Q^* (-\dot{a}h)  
             \Delta^{\beta \gamma}  
     [x^\gamma x^\beta T]  
                          \nonumber \\  
&=& 0.  
\end{eqnarray}  
  
The seventh term of the second moment equation (contraction 2, eq.51) is rewritten by,  
\begin{equation}  
     4 \frac{Gm^2}{a}(ma)\bar{n}^3 a^9 Q^*  
     \Delta^{\beta \gamma}  
     \int   
     \frac{x_{31}^\beta}{x_{31}^3}  
     z^\gamma \{    \frac{\langle v_{31} \rangle}{z}   
                 -  \frac{\langle v_{23} \rangle}{|x-z|}  
               \} \xi_{23}\xi_{31}    
                             d^3 x_3=0.   
\end{equation}  
  
Finally we obtain the following four equations:  
\begin{equation}  
    \frac{\partial \xi}{\partial t}  
+   \frac{1}{a}\frac{1}{x^2}  
\frac{\partial}{\partial x}[x^2(1+\xi)\langle v \rangle]=0,  
~~~~~~~~~~~~~~({\rm 0th~moment})  
\label{0th}  
\end{equation}  
\begin{eqnarray}  
     \frac{1}{a}\frac{\partial}{\partial t}\frac{1}{x^2}  
     \frac{\partial}{\partial x}  
                 [x^2(1+\xi)a\langle v \rangle]  
&+&  \frac{1}{a}\frac{1}{x^2}\frac{\partial}{\partial x}  
       \{\frac{\partial}{\partial x}(x^2(1+\xi) \Pi) - 2x(1+\xi) \Sigma \}  
                               \nonumber \\  
&+&  8\pi Gm\bar{n}a \xi  
                               \nonumber \\  
&+&  2Gm\bar{n}aQM\frac{1}{x^2}\frac{\partial}{\partial x}[x^3 \xi^2]=0,  
                               \nonumber \\  
&&~~~~~~~~~~~~~~~~~~~~~~~~~~~~({\rm 1st~moment})  
\label{1st}  
\end{eqnarray}  
\begin{eqnarray}  
\lefteqn{  
\frac{1}{a^2}\frac{\partial}{\partial t}a^2  
   \frac{1}{x^2}\frac{\partial}{\partial x}  
       \left[  
       \frac{\partial}{\partial x}  
       [x^2(1+\xi)\Pi]-2x(1+\xi) \Sigma   
       \right]  
}  
     \nonumber \\  
&&~~~~+~  \frac{1}{a}\frac{1}{x^2}  
     \frac{\partial^3}{\partial x^3}  
        \left[  
        x^2(1+\xi) \{3\langle v \rangle(\Pi-\Sigma)-2\langle v \rangle^3 \}  
        \right]    
     \nonumber \\  
&&~~~~+~  \frac{1}{a}  
     \frac{3}{x^2}\frac{\partial}{\partial x}  
     \frac{1}{x}\frac{\partial}{\partial x}  
       \left[  
       x^4\frac{\partial}{\partial x}\frac{1}{x}  
       (1+\xi)\langle v \rangle  
       \{\Sigma   +        \frac{2\langle v_1^2 \rangle}{3(1+\xi)} \}  
       \right]  
                                  \nonumber \\  
&&~~~~+~  \frac{1}{a}\frac{1}{x^2}  
     \frac{\partial^3}{\partial x^3}  
     [x^2 (1+\xi)(s_\parallel -3s_\perp)]   
                                   \nonumber \\  
&&~~~~+~  \frac{1}{a}  
     \frac{3}{x^2}\frac{\partial}{\partial x}  
     \frac{1}{x}\frac{\partial}{\partial x}  
       \left[  
       x^4\frac{\partial}{\partial x}\frac{1}{x}  
       (1+\xi)s_\perp  
       \right]  
                                  \nonumber \\  
&&~~~~+~  16 \pi Gm\bar{n}a\frac{1}{x^2}\frac{\partial}{\partial x}  
     x^2 (1+\xi)\langle v \rangle  
                                   \nonumber \\  
&&~~~~+~  4Gm\bar{n}aQ^* (\dot{a}h)\frac{1}{x^2}  
     \frac{\partial^2}{\partial x^2}  
     [x^4 M \xi^2 ] =0,  
                                   \nonumber \\  
&&~~~~  
~~~~~~~~({\rm 2nd~moment:contraction~ 1})  
\label{2nd1}  
\end{eqnarray}  
\begin{eqnarray}  
      \frac{1}{a^2}\frac{\partial}{\partial t}  
      [a^2(1+\xi) \Sigma]   
&+&   \frac{1}{a}\frac{1}{x^4}  
      \frac{\partial}{\partial x}  
          \left[  
           x^4(1+\xi) \langle v \rangle   
           \{\Sigma + \frac{2\langle v_1^2 \rangle}{3(1+\xi)}\}  
           \right]  
                                    \nonumber \\  
&+&   \frac{1}{a}\frac{1}{x^4}  
      \frac{\partial}{\partial x}  
      [x^4 (1+\xi)s_\perp]  
                                    \nonumber \\  
&+&   J=0  
                                    \nonumber \\  
&&~~~~~~~~~~~~~({\rm 2nd~moment~:contraction~ 2})  
\label{2nd2}  
\end{eqnarray}  
where,  
\begin{eqnarray}  
J \equiv  &&\frac{Gm^2}{a}\bar{n}a^3   
          \Delta^{\beta \gamma}  
          \int c(2,3)  
          [   p_{21}^\gamma\frac{x_{31}^\beta}{x_{31}^3}  
           +  p_{21}^\beta\frac{x_{31}^\gamma}{x_{31}^3}]  
          d^3 x_3 d^3 p_3 d^3 p_2                     \nonumber \\  
&&~~~-~   \frac{Gm^2}{a}\bar{n}a^3   
          \Delta^{\beta \gamma}  
          \int c(3,1)  
          [   p_{21}^\gamma\frac{x_{32}^\beta}{x_{32}^3}  
           +  p_{21}^\beta\frac{x_{32}^\gamma}{x_{32}^3}]  
          d^3 x_3 d^3 p_3 d^3 p_1   
\end{eqnarray}  
This term is negligible in the strongly nonlinear regime.  
  In deriving the above equations, we made use of the symmetry of the background universe and assumed that the three-point correlation function can be written by the  products of the two-point correlation functions. And also $\xi$ is assumed to be given by eq.(52) which is expected to be correct in the strongly nonlinear region.

\section{SCALE-INVARIANT SOLUTIONS IN THE STRONGLY NON-LINEAR REGIME}  
  
  In this section, we discuss how the BBGKY equations are approximated   
in the strongly nonlinear limit.  
  And we discuss how the power index of the two-point correlation function   
relates with the skewness, three-point correlation function   
and the mean relative peculiar velocity.  
  Furthermore we investigate whether the stable condition is correct or not.

\subsection{BBGKY Equations in the Nonlinear Limit}  
  
In the strongly nonlinear regime, the two-point correlation function is much   
larger than unity, $\xi \gg 1$ at $x \ll 1$.  
In this limit, the BBGKY equations are given by  
\begin{equation}  
    \frac{\partial \xi}{\partial t}  
+   \frac{1}{a}\frac{1}{x^2}  
\frac{\partial}{\partial x}[x^2 \xi \langle v \rangle]=0,  
~~~~~~({\rm 0th~moment})  
\label{a}  
\end{equation}  
\begin{equation}  
 \frac{1}{a}\frac{1}{x^2}\frac{\partial}{\partial x}  
       \{\frac{\partial}{\partial x}(x^2 \xi \Pi) - 2x \xi \Sigma \}  
+  2Gm\bar{n}aQM\frac{1}{x^2}\frac{\partial}{\partial x}[x^3 \xi^2]=0,  
~~~~({\rm 1st~moment})  
\label{b}  
\end{equation}  
\begin{eqnarray}  
\lefteqn{  
         \frac{1}{a^2}\frac{\partial}{\partial t}a^2  
         \frac{1}{x^2}\frac{\partial}{\partial x}  
         \left[  
                 \frac{\partial}{\partial x}  
                 [x^2 \xi \Pi]-2x \xi \Sigma   
         \right]  
        }     \nonumber \\  
&&~~+  \frac{1}{a}\frac{1}{x^2}  
     \frac{\partial^3}{\partial x^3}  
        \left[  
        x^2 \xi \{3\langle v \rangle(\Pi-\Sigma)+(s_\parallel -3s_\perp) \}  
        \right]    
     \nonumber \\  
&&~~+  \frac{1}{a}  
     \frac{3}{x^2}\frac{\partial}{\partial x}  
     \frac{1}{x}\frac{\partial}{\partial x}  
       \left[  
       x^4\frac{\partial}{\partial x}\frac{1}{x}  
       \xi \{\langle v \rangle \Sigma   + s_\perp  \}  
       \right]  
                                  \nonumber \\  
&&~~-  4Gm\bar{n}aQ^* (\dot{a}h)\frac{1}{x^2}  
     \frac{\partial^2}{\partial x^2}  
     [x^4 M \xi^2 ] =0,  
                                   \nonumber \\  
&&~~~~~~~~~~~~~~~  
~~~~~~~~~~~~~({\rm 2nd~moment:contraction~ 1})  
\label{c}  
\end{eqnarray}  
\begin{eqnarray}  
      \frac{1}{a^2}\frac{\partial}{\partial t}  
      [a^2 \xi \Sigma]   
&+&   \frac{1}{a}\frac{1}{x^4}  
      \frac{\partial}{\partial x}  
          \left[  
           x^4 \xi \{ \langle v \rangle \Sigma + s_\perp \}  
           \right] = 0.  
                                    \nonumber \\  
&&~~~~~~~~~~~~~~  
~~~~~~~~~~~({\rm 2nd~moment:contraction~2})  
\label{d}  
\end{eqnarray}  
  
These equations are equivalent to eqs.(39) $\sim$ (42) in Ruamsuwan \& Fry(1992) while some notations are different from each other.  The zeroth moment equation (63) is derived without any assumption.  Here it must be noted, however, that there is assumption about the three-point correlation function in deriving the first moment equation (64). And in the second moment equations (65) and (66), there are the terms of the skewness and so we usually need the higher moment equations in order to solve eqs.(65) and (66) while  
we do not need them in our analysis.  
  
\subsection{Scale-Invariant Solutions in the Strongly Nonlinear Limit}

In the strongly nonlinear regime, it is naturally expected that the effect of the nonlinear gravitational clustering dominates and then the solutions in this regime have no characteristic scales, that is, they are expected to obey the power law due to the scale-free of the gravity.  So we investigate the power law solutions of the $\xi, \langle v \rangle, \Pi$ and $\Sigma$.  Then we assume that the two-point correlation function $\xi$ is given by  
\begin{equation}  
\xi = \xi_0 a^\beta x^{-\gamma}.  
\end{equation}  
  
We can see from eq.(63) that the mean relative peculiar velocity $\langle v \rangle$ is given by the dimensional analysis as follows:  
\begin{equation}  
\langle v \rangle = -h\dot{a}x  
\end{equation}  
\begin{equation}  
\beta=(3-\gamma)h  
\end{equation}  
  
The stable condition means that $h=1$ because $\langle \dot{r} \rangle  
=\dot{a}x +\langle a\dot{x} \rangle =\dot{a}x+ \langle v \rangle=0$.  
Hereafter we call this parameter $h$, {\it relative velocity parameter}.  
We obtain the power law solutions of the other quantities from eqs.(64) and (65) as follows:  
\begin{eqnarray}  
\Pi&=&\Pi_0 a^{\beta-1}x^{2-\gamma}, \\  
\Sigma&=&\Sigma_0 a^{\beta-1}x^{2-\gamma}\\  
s_\parallel&=&s_{\parallel0}\dot{a} a^{\beta-1}x^{3-\gamma}, \\  
s_\perp&=&s_{\perp0}\dot{a} a^{\beta-1}x^{3-\gamma}.  
\end{eqnarray}  
Here $\Pi_0, \Sigma_0, s_{\parallel0}$ and $s_{\perp0}$ are constants.  
  
  From the above results, we obtain the next relation from eq.(66).  
\begin{equation}  
(1+2\beta)-(7-2\gamma)(h-\Delta)=0.  
\end{equation}  
Here $\Delta$ is defined by  
\begin{equation}  
\Delta \equiv \frac{s_{\perp0}}{\Sigma_0}.  
\end{equation}  
  
We can find from eqs.(69) and (74) that   
\begin{equation}  
h=1+(7-2\gamma)\Delta.  
                           \label{h0}  
\end{equation}  
Here it must be noted that eq.(76) is determined on the assumption that the three-point correlation function can be written by the products of the two-point correlation functions(see eq.[42]).  
Then if this assumption is correct and also the skewness vanishes, which means $\Delta=0$, we can see from eq.(76) that the relative velocity parameter $h$ is equal to unity.  This fact means that the stable condition $h=1$ is {\it not} an assumption, but {\it should be} satisfied in the strongly nonlinear regime when  the three-point correlation function can be represented by the products of the two-point correlation functions and the skewness vanishes.  
However it may be physically natural that the skewness does not vanish in the nonlinear regime while the skewness equals 0 in the linear regime.  Then $h$ should not be equal to 1.  This means that the stable condition is satisfied for the novel case that the skewness vanishes.  Furthermore $h$ varies when the assumption about the three-point correlation function $\zeta$ changes. For example, we assume that $\zeta \propto \xi^{2(1+\delta)}$, here $\delta$ is a constant. Then we can find from eqs.(63) $\sim$ (65) that  
\begin{eqnarray}  
\xi&=&\xi_0 a^{\beta}x^{-\gamma}, \\  
\langle v \rangle &=& -h\dot{a}x,\\  
\Pi&=&\Pi_0 a^{\beta(1+2\delta)-1}x^{2-\gamma(1+2\delta)}, \\  
\Sigma&=&\Sigma_0 a^{\beta(1+2\delta)-1}x^{2-\gamma(1+2\delta)}, \\  
s_\parallel&=&s_{\parallel_0}\dot{a}  
 a^{\beta(1+2\delta)-1}x^{3-\gamma(1+2\delta)}, \\  
s_\perp&=&s_{\perp_0}\dot{a}  
 a^{\beta(1+2\delta)-1}x^{3-\gamma(1+2\delta)}.   
\end{eqnarray}  
  
{}From eq.(66), it is found that  
\begin{equation}  
2\beta(1+\delta)+1-\{7-2\gamma(1+\delta)\}(h-\Delta) = 0,  
\end{equation}  
and then  
\begin{equation}  
h=\frac{1+\{7-2\gamma(1+\delta)\}\Delta}{1-6\delta}   
                                    \label{h}.  
\end{equation}  
  
Thus we can see that the value of the relative velocity parameter $h$ also depends on the assumption about the three-point correlation function.  
It must be noted here that a little change of $\delta$ results in the large change of $h$ due to the factor $(1-6\delta)$ in eq.(84).

\subsection{Relative Velocity Parameter}  
  
In the previous subsection, we found that the relative velocity parameter $h$ depends on the three-point correlation function and the skewness.  Then, in this subsection, we discuss about the allowed value of $h$ from the physical point of view.  
  
In general, the mean relative  velocity $\langle \dot{r} \rangle$ is represented as follows:   
\begin{equation}  
\langle \dot{r} \rangle=\langle \dot{a}x + a \dot{x} \rangle,  
\end{equation}  
and then the mean relative peculiar velocity $\langle v \rangle$ is given by 
\begin{eqnarray}  
\langle v \rangle &=&  \langle a \dot{x} \rangle \nonumber \\  
                  &=&   -\dot{a}x +\langle \dot{r} \rangle \nonumber \\  
             &\equiv&   -h\dot{a}x.  
\end{eqnarray}  
  
DP assumed the stable condition in which $h=1$ in the virialized objects.  But we found that $h$ depends on the $\xi$ and  $\Delta$. The stable condition is satisfied only in the novel case that $\Delta=0$ if eq.(42) is correct.  
We show here that even in the virialized region $h$ can have other values rather than unity.  For example, we consider the spherical collapse for simplicity.  In this case, it is found that a object has collapsed and virialized when the linearly estimated density fluctuation $\delta_{lin}$ of this object becomes just 1.69($\equiv \delta_c$)(Peebles 1993).  When   
the $\delta_{lin}$ of the virialized object with mass scale $M$ becomes greater than  $\delta_c$, then the object with mass scale $M'>M$ should collapse and virialized at the same place(Press \& Schechter 1974; Yano, Ngashima \& Gouda 1996).  
This means physically that matters around the virialized object accrete  
to this object with mass $M$ and the larger virialized object with mass $M'$ are newly formed.  
In this situation, the linearly estimated  density fluctuations of the virialized object is given by $\delta_{lin}=\delta_c$, which means the real density fluctuations $\rho_{nonlin}=18 \pi^2 \langle \rho \rangle$(Suto, 1993). Here $\langle \rho \rangle$ is the mean density of the universe at the collapsed time. Thus the density of the virialized object $\rho_{nonlin}$  is in proportion to $\langle \rho \rangle \propto a^{-3}_c$,  
where $a_c$ is the scale factor when the object has just virialized.  
  
Then we consider a sphere of the virialized object with physical radius $r$ and consider the change of the mass of the shell at $r$.  The mass change is represented as follows:  
\begin{eqnarray}  
4\pi r^2 \rho \langle \dot{r} \rangle   
               &=& -\frac{4}{3}\pi r^3 \frac{d\rho}{dt} \nonumber \\  
               &=& -\frac{4}{3}\pi r^3 (-3)\frac{\dot{a}}{a}\rho.   
\end{eqnarray}  
  
Together with $r=ax$, we obtain  
\begin{equation}  
\langle \dot{r} \rangle   = \dot{a}x.   
\end{equation}  
  
And then the mean relative peculiar velocity is given by  
\begin{eqnarray}  
\langle v \rangle &=&   -\dot{a}x +\langle \dot{r} \rangle \nonumber \\  
                  &=& 0  
\end{eqnarray}  
  
In this case, the relative velocity parameter $h$ is equal to 0.  
This case is one example of the physical situation that $h$ is not equal to 1 even in the virialized regions.  
  
In the hierarchical clustering picture, the variance of density fluctuations on small scales is larger than that on the larger mass scale and the smaller objects are formed at first and after that the larger objects are formed.  
In this case, we can consider two extreme physical situations in the process that the larger objects are formed.  One of them is as follows; the smaller collapsed objects have clustered and survived in the larger object.  In this situation, the objects which are formed are never broken and the mean interval of particles in the objects does not change.  Hence the relative velocity parameter $h$ is equal to 1.  Another situation is as follows; matters around the virialized object accrete to this object and the large virialized object are newly formed.  In this situation, the mean interval of particles expand with the Hubble velocity as shown in eq.(88).  
  
{}From the above argument, we can conclude that the way of clustering decides the value of the relative velocity parameter $h$.  We believe that the real way of the clustering might be between the above two extreme situations.  The relative velocity paraemter $h$ should not be greater than 1 because $\langle \dot{r} \rangle$ have to be positive. And as we mentioned above, $h=0$ is the lower limit of $h$ for the virialized object.  
Thus the relative parameter $h$ takes the value between 0 and 1,  
\begin{equation}  
0 \leq h \leq 1.  
\end{equation}  
This condition gives constraint on the skewness $\Delta$ and the three-point correlation function $\zeta$.

\subsection{Self-Similar Solutions}

DP showed the existence of the self-similar solutions under the stable condition and the assumptions about the three-point correlation function and the skewness.  In the self-similar solutions, the power index of the two-point correlation function in the strongly nonlinear regime is related to the power index $n$ of the initial power spectrum, where $n$ is defined by  
\begin{equation}  
 P(k) \propto k^n.   
\end{equation}  
  
In the linear regime, the two-point correlation function is given by(Peebles 1980,1993),  
\begin{equation}  
\xi \propto a^2 x^{-(3+n)} \propto (\frac{x}{a^\alpha})^{-(3+n)}  
\end{equation}  
where  
\begin{equation}  
\alpha \equiv \frac{2}{3+n}.  
\end{equation}  
  
On the other hand,   
in the strongly nonlinear regime, the two-point correlation function obeys the following evolution equation(0th moment, eq.[63]),  
\begin{equation}  
   a \frac{\partial \xi}{\partial a}  
-  h \frac{1}{x^2}  
\frac{\partial}{\partial x}[x^3 \xi]=0.  
\end{equation}  
Here we used the relation $\langle v \rangle=-ha \dot{x}$.  
  
This equation is rewritten by transforming the variables, $x$ and $a$ to   
the scaling variable, $s \equiv x/a^{\beta}$,   
\begin{equation}  
\frac{d\xi}{ds}=-\frac{3h}{\beta +h}\frac{\xi}{s}  
\end{equation}

If the self-similarity is satisfied for the scaling variable $s$, $\beta$ should be equal to $\alpha={2}/(3+n)$.  
  
Then it is found from eq.(95) that   
\begin{equation}  
\xi \propto s^{-\frac{3h}{\alpha +h}}=s^{-\gamma},  
\end{equation}  
and the power index $\gamma$ is given by  
\begin{equation}  
\gamma= \frac{3h}{\alpha +h}=\frac{3h(3+n)}{2+h(n+3)}.  
                              \label{gammah}  
\end{equation}  
  
  If the relative velocity parameter $h=1$ is satisfied,    
$\gamma=3(3+n)/(5+n)$, which is the result shown by DP.  
   
In the previous subsection, the allowed range of $h$ is shown($ 0 \leq h \leq 1$).  Then we get the range of the power index $\gamma$ as follows:  
\begin{equation}  
0 \leq \gamma \leq \frac{3(3+n)}{2+(n+3)}.  
\end{equation}  
  
Now we investigate the possibility that there exist the self-similar solutions in which the power index $\gamma$ is independent of the initial power index $n$. We can see from eq.(97) that this situation occurs when the relative velocity parameter $h$ satisfies the following relation  
(Padmanabhan(1995)),  
\begin{equation}  
h= \frac{c}{3+n},  
                              \label{h-n}  
\end{equation}  
where $c$ is a constant.  
  
However $h$ should be within the range between 0 and 1.  If $c$ is not equal to 0, $h$ becomes the value of larger than unity for some $n$ even if $h$ is within the allowed region for other $n$.  
Hence only when $c$ is equal to 0, that is, $h=0$, we can get the self-similar solution in which the power index $\gamma$ is irrespective of $n$ and also $h$ is always within the allowed region. This result is reasonable from the physical point of view shown in $\S 3.3$. When $h=0$, the smaller collapsed objects become to be absorbed hierarchically in the larger object. In this case, the information of the initial memories are erased.  Then the power index $\gamma$ is irrespective of $n$.  On the other hand, when $h \neq 0$, substructures in the larger cluster survived and the initial memories are not completely erased.

\section{RESULTS AND DISCUSSION}  
  
In this paper, we investigated the scale-invariant solutions of the cosmological BBGKY equations in the strongly nonlinear regime. The mean relative peculiar  
velocity depends on the skewness and the three-point correlation function. The stable condition which DP used is satisfied for the novel case that the skewness vanishes and the three-point correlation function can be represented by the products of the two-point correlation functions.  Furthermore the power index of the two-point correlation function in the strongly nonlinear regime depends on $h$, the skewness and the three-point correlation function(see eqs.[76] and [84]).  
  
  In the hierarchical clustering, there are two extreme situations in the way of the clustering.  One of them is that the collapsed object can not be broken and clustered together to form the larger cluster.  In this situation, the mean separation of the particles does not change as time increases and $h=1$.  This corresponds to the stable condition.  Another situation is that matters around the collapsed object accrete to this object and the larger collapsed object is newly formed.  In this case, the mean separation of the particles is expanding with the Hubble velocity and then $h=0$.  It is believed that in general $h$ should take the value between 0 and 1.  This constrains the skewness and the three-point correlation function.  In general, the skewness does not vanish in the nonlinear regime and so it might be impossible for the stable condition$(h=1)$ to be satisfied.

The self-similar solution is shown by DP under some assumptions and the stable condition.  If there exist the self-similar solutions under the other conditions, then the power index $\gamma$ is given by  
\begin{equation}  
0 \leq \gamma \leq \frac{3(3+n)}{5+n},  
\end{equation}  
 because $0 \le h \le 1$.  
  
In general we may expect that the power index $\gamma$ does not depend on $n$ because the systems forget the initial memories on the nonlinear regions due to the nonlinearity of the gravity. If $h$ equals $c/(n+3)$, then $\gamma$ is irrespective of $n$ and is given by $3c/(2+c)$. 
  However $h$ should be between 0 and 1. Then $c$ should be zero, that is, $h=0$. 
  In this case, as mentioned before, the collapsed objects are hierarchically absorbed into the larger objects and the substructures in the cluster can not survive.  Then the initial memories are erased.  
Thus it is physically reasonable that $\gamma$ does not depend on the initial condition $n$ in this case.

In general, the systems have substructures in some clusters and no subclusters in other clusters and so $\gamma$ depends on $n$.  However the relation between $\gamma$ and $n$ may not be given by the result shown by DP because the skewness does not vanish in general on the nonlinear regions and so $h$ should not be equal to 1. Here we comment that $\gamma=0$ is expected from the catastrophe theory.  And we can see that if the solutions of the correlation function have the self-similarity, the power index $\gamma=2$, which Saslaw(1980) predicted,  is impossible to appear stably in the strongly nonlinear regime irrespectively of the initial power index $n$ because in this case $c$ should be eqaul to 4 and.then $h>1$ for $n<1$.  
  
Ruamsuwan \& Fry tested the stability of the scale-invariant solutions derived by DP and found that they are marginally stable.  We will analyze the instability of the scale-invariant solutions shown in this paper under the other assumptions beside those taken in DP. Furthermore we will investigate whether there exist the self-similar solutions under the various assumptions about the three-point correlation function and the skewness in the future.  
   
\acknowledgements  
  
We are grateful to M.Nagashima for useful discussions. And we would like to thank S.Ikeuchi and M.Sasaki for useful suggestions and continuous encouragement.  This work was supported in part by the Grant-in-Aid No.06640352 for the Scientific Research Fund from the Ministry of Education, Science and Culture of Japan.

\end{document}